\documentclass{midl} 
\usepackage{mathtools}
\newcommand\independent{\protect\mathpalette{\protect\independenT}{\perp}}
\def\independenT#1#2{\mathrel{\rlap{$#1#2$}\mkern2mu{#1#2}}}

\DeclareMathOperator*{\argmin}{arg\,min}
\usepackage{mwe}

\jmlryear{2022}
\jmlrworkshop{Full Paper -- MIDL 2022}
\title[Personalized Treatment Using MRI]{Personalized Prediction of Future Lesion Activity and Treatment Effect in Multiple Sclerosis from Baseline MRI}

\midlauthor{\Name{Joshua Durso-Finley\nametag{$^{1}$}} \Email{jddursof@cim.mcgill.ca}\\
\Name{Jean-Pierre R. Falet\nametag{$^{1,2}$}} \Email{jpfalet@cim.mcgill.ca}\\
\Name{Brennan Nichyporuk\nametag{$^{1}$}} \Email{brennann@cim.mcgill.ca}\\
\Name{Douglas L. Arnold\nametag{$^{2}$}} \Email{douglas.arnold@mcgill.ca}\\
\Name{Tal Arbel\nametag{$^{1}$}} \Email{arbel@cim.mcgill.ca}\\
\addr $^{1}$ Centre for Intelligent Machines, Department of Electrical and Computer Engineering, McGill University \& MILA Quebec AI Institute, Canada \\
\addr $^{2}$ Montreal Neurological Institute, McGill University, Montreal, Canada
}

\begin{document}
\maketitle
\begin{abstract}
Precision medicine for chronic diseases such as multiple sclerosis (MS) involves choosing a treatment which best balances efficacy and side effects/preferences for individual patients. Making this choice as early as possible is important, as delays in finding an effective therapy can lead to irreversible disability accrual. To this end, we present the first deep neural network model for individualized treatment decisions from baseline magnetic resonance imaging (MRI) (with clinical information if available) for MS patients. Our model (a) predicts future new and enlarging T2 weighted (NE-T2) lesion counts on follow-up MRI on multiple treatments and (b) estimates the conditional average treatment effect (CATE), as defined by the predicted future suppression of NE-T2 lesions, between different treatment options relative to placebo. Our model is validated on a proprietary federated dataset of 1817 multi-sequence MRIs acquired from MS patients during four multi-centre randomized clinical trials. Our framework achieves high average precision in the binarized regression of future NE-T2 lesions on five different treatments, identifies heterogeneous treatment effects, and provides a personalized treatment recommendation that accounts for treatment-associated risk (e.g. side effects, patient preference, administration difficulties).

\end{abstract}


\section{Introduction}
Precision medicine involves choosing a treatment that best balances efficacy against side effects/personal preference for the individual. In many clinical contexts, delays in finding an effective treatment can lead to significant morbidity and irreversible disability accrual. Such is the case for multiple sclerosis, a chronic neurological disease of the central nervous system. Although numerous treatments are available, each has a different efficacy and risk profile, complicating the task of choosing the optimal treatment for a particular patient. One hallmark of MS is the appearance of lesions visible on T2-weighted MRI sequences of the brain and spinal cord~\cite{RRMSactivity}. The appearance of new or enlarging, NE-T2, lesions on sequential MRI indicates new disease activity. Suppression of NE-T2 lesions constitutes a surrogate outcome used to measure treatment efficacy. Predicting the future effect of a treatments on NE-T2 lesions counts using brain MRI prior to treatment initiation would therefore have the potential to be an early and non-invasive mechanism to significantly improve patient outcomes.
 
Predicting future treatment effects first requires accurate prognostic models for future disease evolution. Deep learning has been used to predict prognostic outcomes in a variety of medical imaging domains~\cite{SmokingPrognosis,StrokePrognosis,AlzeheimersPrognosis,BRATSSurvivalprog}. In the context of MS, research has mainly focused on the related tasks of lesion segmentation~\cite{VALVERDE2017159MSseg,MSSEGFULLYCONV,TanyaUncMSseg,nichyporuk2021cohort} and NE-T2 lesion detection~\cite{DoyleLesionDetection,nazactivitypred}. Recently, deep learning models have been developed for the binary prediction of future disability progression~\cite{pmlr-v102-tousignant19a} and the binary prediction of future lesion activity~\cite{NazFirstActivityPaper}, as defined by the presence of more than one NE-T2 or Gadolinium enhancing lesions. The prediction of more granular outcomes, such as future NE-T2 lesion counts, remains an open research topic. Furthermore, models are typically built as prognostic models for untreated patients. Predicting prognosis on treatment requires addressing the additional challenge of learning the effect each treatment will have on a particular patient based on their MRI, and thus potentially subtle MRI markers predictive of future treatment response.  Machine learning models that have been devised to predict treatment response when it is directly measurable on the image (e.g. shrinking tumour)~\cite{XuLung58,TreatmentEffectBreastTumor}, are insufficient for the context of MS and for other diseases where treatment response must be evaluated relative to placebo or other treatments. Previous work by~\cite{Doyle2017PredictingFD} examined the ability of classical machine learning models to perform binary activity prediction for patients on MS treatments and identify potential treatment responders. 

Several machine learning methods have been developed to estimate treatment effects for single treatment-control comparisons \cite{louizos2017causal, shalit2017TARNET, shi2019adaptingDRAGONNET}, with extensions to multiple treatments \cite{ZhaoCTS, zhao2020mutli-treatment}. \citet{zhao2020mutli-treatment} also integrate the notion of \textit{value} and \textit{cost} (or risk) associated with a treatment, crucial elements for making sound recommendations, particularly when higher efficacy medications may be associated with more severe side effects. However, applications to precision medicine have largely focused on using clinical data as input \cite{DeepSurv,Fotso2018DeepNN, Ching2018CoxnnetAA, Jaroszewicz2014UpliftMW}. Existing MS models \cite{MPScoringCoxPH, RioScore, OneYearMSINF-B} are also limited to  clinical features (e.g. demographics), and established group-level MRI-\textit{derived} features (e.g. contrast-enhancing lesion counts, brain volume). Deep learning models would permit learning individual, data-driven features of treatment effect directly from MRI sequences and should provide improvement on existing strategies.

This paper introduces the first image-based treatment recommendation framework for MS that combines prognosis prediction, treatment effect estimation, and treatment-associated risk (\figureref{fig:block_diagram}) evaluation. Our models takes multi-sequence MRI at baseline, along with available clinical information, as input to a multi-head deep neural network that learns shared latent features in a common ResNet encoder~\cite{he2015deepResNet}. It then learns treatment-specific latent features in each output head for predicting future potential outcomes on multiple treatments. Predictions, effect estimates, and treatment risk are then supplied to a {\it Clinical Decision Support Tool} that outputs a treatment recommendation. This framework is evaluated on a proprietary multi-trial, multi-scanner dataset of MS patients exposed to five different treatment options. The multi-head model not only accurately predicts, from baseline, future NE-T2 lesion counts that will develop 1-2 years ahead on all treatments, but it is able to reliably identify subgroups with heterogeneous treatment effects (groups for which the treatment is more or less effective) as measured by causal inference metrics. Finally, this framework shows that improved lesion suppression can be achieved using the support tool, especially when treatment risk is being considered.

\begin{figure}
  \floatconts
  {fig:block_diagram}
    {\caption{System overview illustrating the overall approach.}} 
  {\centering
  \includegraphics[width=0.95\textwidth]{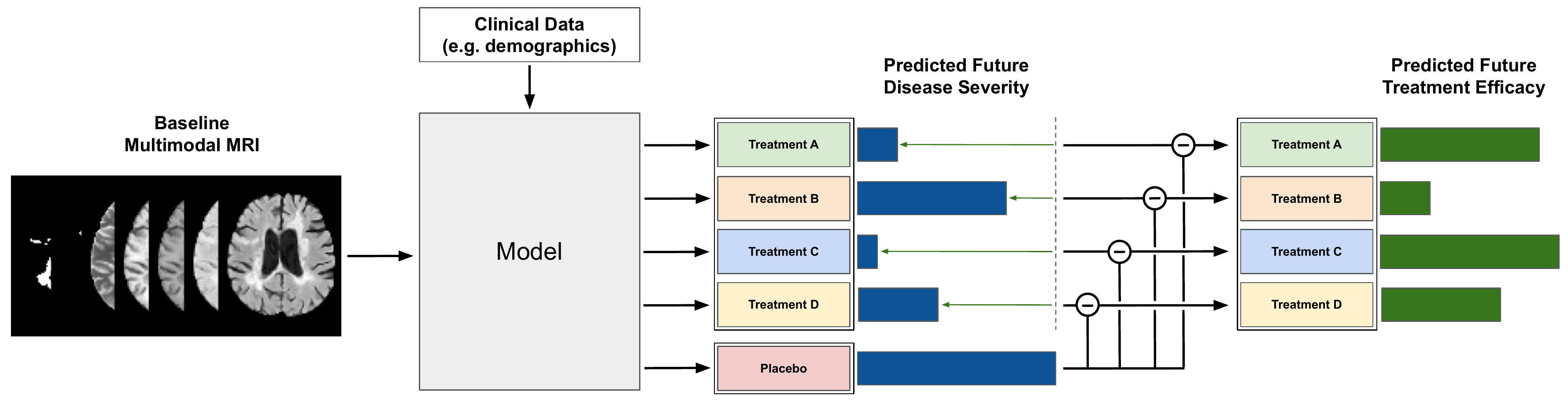}}
\end{figure}

\section{Method}
\subsection{Estimating Treatment Effect}
Let $X \in \mathbb{R}^d$ be the input features (multi-sequence MRI and available clinical data), $Y \in \mathbb{R}$ be the outcome of interest, and $W \in \{0, 1, ..., m\}$ be the treatment allocation in the case where $w=0$ is a control (e.g. placebo) and the remaining are $m$ treatment options. Given an observational dataset $\mathcal{D} = \{(x_i, y_i, w_i)\}_{i=1}^n$, 
the individual treatment effect (ITE) for patient $i$ can be defined using the Neyman/Rubin Potential Outcome Framework \cite{Rubin1974EstimatingCE} as $ITE_i=Y_i(t) - Y_i(0)$, 
where $Y_i(t)$ and $Y_i(0)$ represents \textit{potential} outcomes on treatment $t \in \{1, ..., m\}$ and control, respectively. The ITE is therefore a fundamentally unobservable causal quantity because only one of these potential outcomes is realized. Treatment effect estimation in machine learning therefore relies on a related causal estimand, the conditional average treatment effect (CATE)

\begin{equation}
    \tau_t(x) = \mathbb{E}[Y(t)|X=x] - \mathbb{E}[Y(0)|X=x].
\end{equation}

The causal expectations can be recovered from the observational data as follows
\begin{equation}
    \tau_t(x) = \mathbb{E}[Y|X=x, W=t] - \mathbb{E}[Y|X=x, W=0] = \mu_t(x) - \mu_0(x)
\end{equation}
which can be estimated in an unbiased fashion using randomized control trial data (as in our case), where $\{(Y(0), Y(1))\}$ $\independent W|X$ \cite{Gutierrez2017}. Further assumptions are needed in the context of non-randomized data \cite{PropensityScoreUsage}.

\subsection{Network Architecture}
\label{sec:net}
\begin{figure}[htbp]
 \floatconts
   {fig:NetworkDiagram}
   {\caption{Network Diagram. Common ResNet encoder followed by treatment-specific output MLPs for predicting potential outcomes on multiple treatments.}}
   {\includegraphics[width=.95\linewidth]{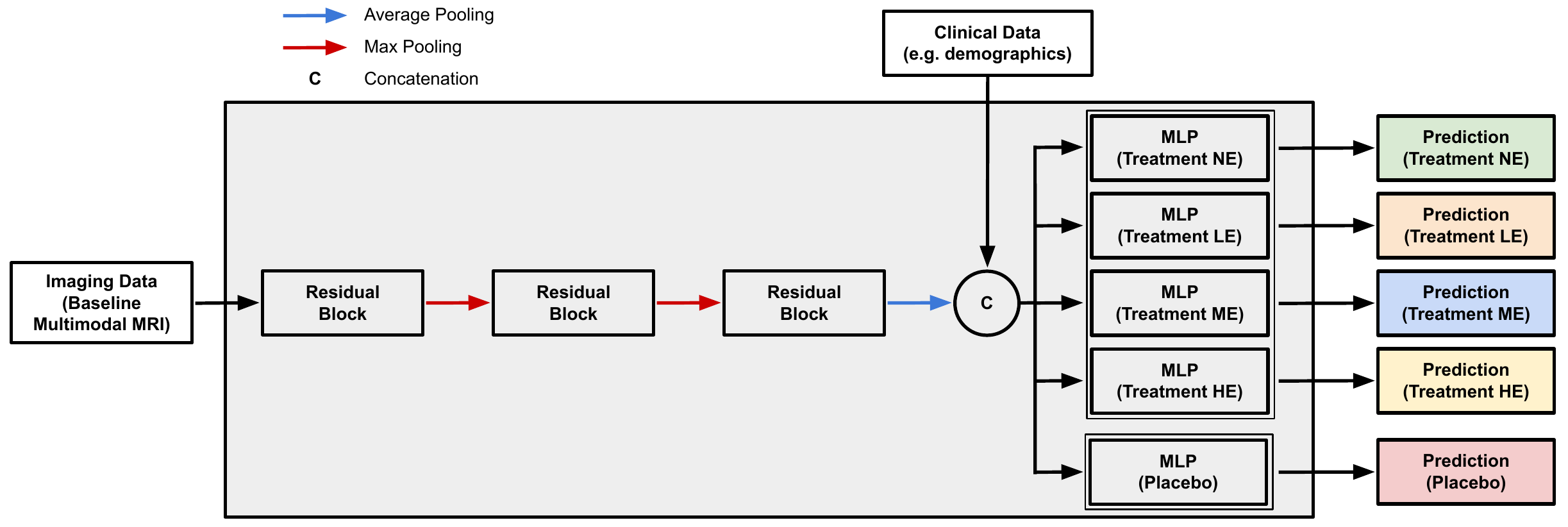}}
   \label{fig:NetworkDiagram}
 \end{figure}

Our network is based on TARNET \cite{shalit2017TARNET} and its multi-treatment extension \cite{zhao2020mutli-treatment}. Specifically, we employ a single multi-head neural network composed of $m$ different CATE estimators,
\begin{equation}
    \hat{\tau}_{t}(x) = \hat{\mu}_t(x) - \hat{\mu}_0(x),\ t\in\{1,...,m\}
\end{equation}
where each $\hat{\mu}_t(x)$ is parametrized by a neural network trained on the corresponding treatment distribution, and all share parameters in the earlier layers. A ResNet encoder is used as the shared trunk, and after a global max pooling layer, the encoded features are concatenated with any available clinical information before being processed by treatment-specific multilayer perceptrons (MLPs). The model architecture is depicted in \figureref{fig:NetworkDiagram}. 

During training, mini-batches are randomly sampled from $\mathcal{D}$ and fed through the network, outputting a prediction for each treatment head. Losses are computed at each head $t$ for the set of prediction-target pairs where ground truth is available for that treatment, $\{(\hat{y}_{i,t}, y_i)\}_{i:w_i=t}$. Shared parameters are learned in the common layers, which receive gradients for each sample irrespective of treatment allocation, while treatment-specific parameters are learned in the treatment heads from samples allocated to the corresponding treatment. At inference, predictions from all output heads are used for every patient. Full implementation details can be seen in \appendixref{sec:ImpDets}.

The tasks of regression and classification are examined. Regressing future NE-T2 lesion counts offers the most intuitive interpretation of treatment effect $\hat{\tau}_{t}(x)$ (i.e. differences in lesion count), but is sensitive to outliers in the count distribution (e.g. patients with 50 lesions). On the other hand, MS guidelines \cite{MSguidelines} report a cutoff of ($\geq 3$) new/enlarging T2 lesions after which a treatment should be changed to a more effective one.  We therefore also consider the binary classification task of predicting minimal evidence of disease activity on future T2 sequences, referred to as {\it MEDA-T2}, as having $<3$ NE-T2 lesions. Unfortunately, the treatment effect $\hat{\tau}_{t}(x)$ at the binary scale would not capture the true range of effects, and using the softmax outputs to compute $\hat{\tau}_{t}(x)$ has a less informative interpretation as compared to regressed counts. For the regression loss, we use Mean Squared Error (MSE) on the log-transformed count, $ln(y_i+1)$, to reduce the weight of outliers. For the classification loss, we use binary cross entropy (BCE) on the binary MEDA-T2 outcome, $I(y_i<3)$, where $I(\cdot)$ is the indicator function.

\subsection{Clinical Decision Support Tool}
\label{sec:ptp}
Based on \citet{zhao2020mutli-treatment}, 
we define $r_t$ to be the risk associated with treatment $t \in \{1, 2, ..., m\}$. This can be set by a clinician and patient based on their experience/preference, or could be extrapolated from long-term drug safety data. In the case of MS, drugs can be grouped into lower efficacy (LE), moderate efficacy (ME), and high efficacy (HE). An escalation strategy (starting with LE and escalating if necessary) is often used to avoid unnecessarily exposing patients to side effects attributed to higher efficacy drugs \cite{msescalationguidelines}. We therefore set $r_t = c_t\lambda$, where $\lambda$ is the constant incremental risk associated with moving up the ladder of efficacy (which is set by the user). $c_t$ takes on a value of $0$ for placebo, $1$ for LE, $2$ for ME, and $3$ for HE. We define risk-adjusted CATE, as
\begin{equation}
    \hat{\tau}_t^*(x) = \hat{\tau}_t(x) + r_t.
\end{equation}
Assuming negative CATE indicates benefit, here a reduction in NE-T2 lesions, the tool
then recommends treatment $j$ such that $ j=\argmin_t \hat{\tau}_t^*(x).$

\section{Experiments and Results}
\label{sec:exp}
\subsection{Dataset}
The dataset is composed of patients from four randomized clinical trials: BRAVO \cite{BRAVO}, OPERA 1 \cite{OPERA}, OPERA 2 \cite{OPERA}, and DEFINE \cite{DEFINE}. Each trial enrolled patients with relapsing-remitting MS (the most common form) and had similar recruitment criteria.  We excluded patients who did not complete all required MRI timepoints, or were missing MRI sequences/clinical features at baseline, resulting in a dataset with $n=1817$. Treatments for these trials are categorized based on their efficacy at the group level: placebo ($n=362$), no efficacy (NE, $n=261$), lower efficacy (LE, $n=295$), moderate efficacy (ME, $n=431$), and high efficacy (HE, $n=468$) with each level representing one treatment. Pre-trial statistics and treatment distributions can be seen in Appendix \ref{sec:PretrialStats}.

All trials acquired MRIs at 1 x 1 x 3 mm resolution at the following timepoints: baseline (prior to treatment initiation), one year, and two years. Each contains 5 sequences: T1-weighted, T1-weighted with gadolinium contrast agent, T2-weighted, Fluid Attenuated Inverse Recovery, and Proton Density weighted. In addition, expert-annotated  gadolinium-enhancing (Gad) lesion masks and T2 lesion labels are provided. The baseline MRIs and lesion masks were used as input to our model, while the NE-T2 lesion counts occurring between year one and two were used to compute count target and the binarized MEDA-T2 outcome. Patient's who did not complete all the required MRIs were excluded as they would not have a NE-T2 count. Percentage of MEDA-T2 in our dataset for placebo, NE, LE, ME, and HE are is 45.7\%, 54.4\%, 63.8\%, 77.4\%, 99.6\%, respectively. In addition, baseline age, sex, and Expanded Disabillity Status Scale \cite{Kurtzke1444EDSS}, a clinical disability score, were used as additional clinical features as inputs to our model. 
The dataset was divided into a 4x4 nested cross validation scheme for model evaluation \cite{NestedCrossVal}. Following \citet{Soltys2014EnsembleMF}'s use of ensembling, the 4 inner-fold models are used as members of an ensemble whose prediction on the outer fold's test set is the average of its members.

\subsection{Predicting Future Lesion Suppression}
\label{sec:fact_exp}
We conduct three experiments to determine the best performing framework for predicting the observed future MEDA-T2 given different combinations of inputs, targets, and loss functions. The first compares the performance of the proposed single multi-head architecture with the performance of $m$ independently trained networks. The second assesses the benefit of using both imaging and clinical features. The third compares binary classification of MEDA-T2 with binarization of the output of a regression model trained directly on the NE-T2 lesion counts. 
\begin{table}[htbp]
    \floatconts
    {tab:factualPR}
    {\caption{Average precision scores for the binary MEDA-T2 outcome.}}
     {\resizebox{.8\columnwidth}{!}{\begin{tabular}{|c|c|c|c|c|c|c|c|}
     \hline
     Model Type & $+$ Clinical & Multi-Head & Placebo AP  &  NE AP &  LE AP &   ME AP  &  HE AP  \\
     \hline
     Random Baseline & & & 0.457 &  0.544 & 0.638 & 0.774 & \textbf{0.996} \\
     \hline
     Clinical Only & \checkmark & & 0.72 +/- .08 &  0.76 +/- .02 & 0.82 +/- .06 & 0.90 +/- .03 & 0.995 +/- .01 \\
     \hline
     Binary Classification& \checkmark &  & 0.78 +/- .04&  0.76 +/- .06 & 0.79+/-.03 & {\bf 0.916}+/-.02 & 0.997 +/- 0.01 \\
     \hline
     Binary  Classification & & \checkmark & 0.71 +/-.09 &  0.70 +/-.01 & 0.82 +/-.05 & 0.9 +/-.01 & 0.995 +/-.01 \\
     \hline
     Binary Classification & \checkmark & \checkmark & 0.78 +/-.08 &  0.79 +/-.03 & 0.86+/-.04 & 0.9 +/-.04 & 0.995 +/-.01 \\
     \hline

     Binarized Regression & \checkmark & \checkmark & {\bf 0.80} +/- .08 &  {\bf 0.79}+/- .01 & {\bf 0.87} +/- .04 & 0.913+/- .03 & 0.996 +/-.01 \\
     \hline
\end{tabular}}}
\end{table}
%

\begin{table}[htbp]
    \floatconts
    {tab:factualRegressionTarget}
    {\caption{MSE for log lesion count regression against baseline (mean log lesion count).}}
     {\resizebox{.45\columnwidth}{!}{\begin{tabular}{|c|c|c|c|c|c|}
     \hline
      Model &  Placebo  &  NE &  LE &   ME  &  HE  \\
      \hline
       Baseline &  1.273 &  1.311 & 1.0432 &  0.904 & 0.0443 \\
     \hline
      Regression & 0.669 & 1.062 & 0.849 & 0.701 &0.0433 \\
      \hline
\end{tabular}}
}
\end{table}
Model performance is evaluated using average precision (AP) due to class imbalances in some of the treatment arms, particularly on HE. The random baseline reflects the positive MEDA-T2 label fraction on each arm. For an improved estimate of the generalization error, metrics are computed from the aggregated outer fold test set predictions. Results are shown in \tableref{tab:factualPR}. The multi-head architecture improves APs across most treatment arms, and the concatenation of clinical features provides an additional boost in performance. Finally, the multi-head binarized regression model with clinical data concatenation outperformed the binary classification equivalent. 

Given its strong performance, we  performed the following evaluations using the regression model. We evaluated the MSE on the non-binarized output of the regression model (the log-lesion count), which demonstrates an improvement over the random baseline (mean log lesion count) for all treatments except HE (see \tableref{tab:factualRegressionTarget}). The failure to regress lesion counts on HE can be explained by the extremely small variance in the target distribution, with only 5\% of all test patients having $>0$ future NE-T2 lesion counts.

\subsection{Estimating Treatment Effects}
Given that the regression model outperforms alternatives on MEDA-T2 classification, and because it provides added granularity and a more intuitive interpretation, we  used this model for CATE estimation. 
CATE estimates are computed for each treatment arm relative to placebo.

\begin{figure}[htbp]
\floatconts
  {fig:ADNEWT2BIN}
  {\caption{Treatment Effect Analysis. (a) Average lesion count differences between treatment-placebo pairs, binned according to tertiles of predicted treatment effect size. P-values for differences between groups are shown in Appendix \ref{sec:SignificanceValues}. (b) Average risk-adjusted lesion count for individuals who did (blue) or did not (orange) receive the recommended treatment, compared to random treatment assignment (green). Incremental risk values ($\lambda$) are varied on the $x$-axis.}}
  {
    \subfigure{
    \centering
    \label{subfig:countAD}
    \includegraphics[width=.45\linewidth]{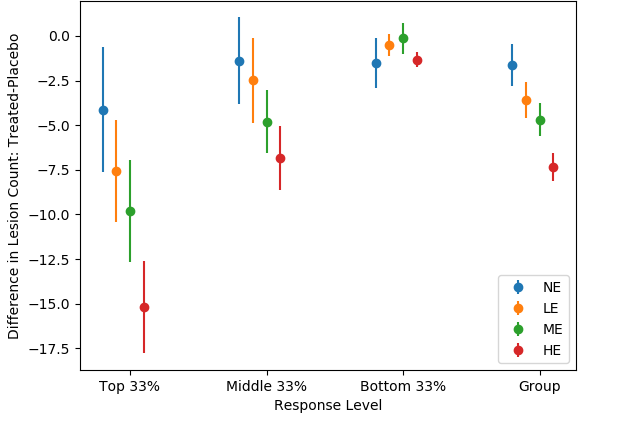}}
      \subfigure{
      \label{subfig:countCDST}
      \centering
    \includegraphics[width=.45\linewidth]{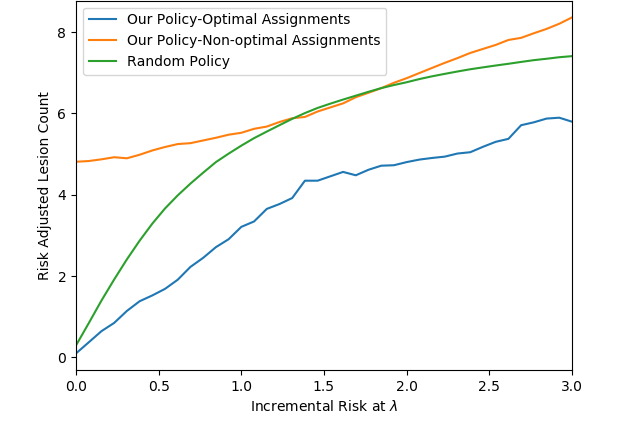}}}
\end{figure}

To evaluate the quality of the CATE estimation, we report uplift bins \cite{AscarzeQuartileAD} at three thresholds of predicted effect. Response ($\hat{\tau}_t$) values are binned into tertiles, and the average difference between the ground truth lesion count for patients who factually received the treatment $t$ and those who factually received placebo is computed for each treatment $t$. The result, shown in \figureref{subfig:countAD}, demonstrates individuals predicted to respond most (top 33\%) have a significantly greater reduction in lesion count over the entire group, and the ones predicted to respond least (bottom 33\%) have a smaller reduction than the entire group. This suggests the model correctly identifies heterogeneous treatment effects. Uplift bins at different resolutions can be seen in Appendix \ref{sec:ADquindec}.

\subsection{Clinical Decision Support Tool In Action}
We now illustrate how the tool could be used in practice. Assuming each drug is associated with a different risk profile (see \sectionref{sec:ptp}), \figureref{fig:Examplepatients} illustrates examples of potential outcomes for two patients. Patient (a) might opt for either a HE efficacy option if they are not worried about greater risk of side effects or cost, or might select a ME option if they are more risk-averse. Patient (b), in turn, might opt for a drug that is NE at the group level but that is predicted to be of comparable efficacy to other options in their particular case.

Individual potential outcome predictions cannot be evaluated due to the lack of ground truth, but we can evaluate the group outcomes for those who received the recommended treatment. To do so, we adjust the ground-truth future NE-T2 lesion count for each individual who received the recommended treatment by adding the risk associated with that treatment,  $y_i^* = y_i + r_t$, and compare their average risk-adjusted lesion count to the group who received a non-recommended treatment (\figureref{subfig:countCDST}). Patients who were factually assigned treatment based on the system's recommendation had a reduction in expected adjusted lesion count for any value of the incremental cost $\lambda$ (varied along the $x$-axis) which indicates the tool provides better treatment recommendations when minimizing treatment-associated risk.

\begin{figure}[htpb]
    \floatconts
    {fig:Examplepatients}
    {\caption{Predicted future lesion count on each treatment for two different test patients. Error bars indicate the standard deviation of the ensemble prediction.The MEDA-T2 threshold (3 lesions) is depicted by the dashed line.}}
    {
    \centering
    \subfigure{
        \centering
        \includegraphics[width=.45\linewidth]{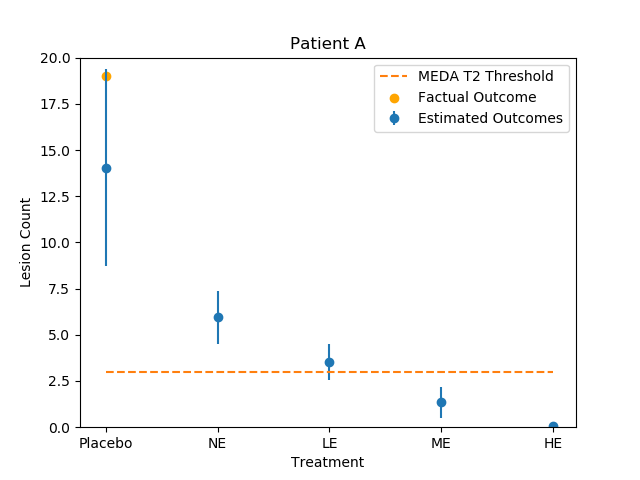}}
    \subfigure{
        \centering
        \includegraphics[width=.45\linewidth]{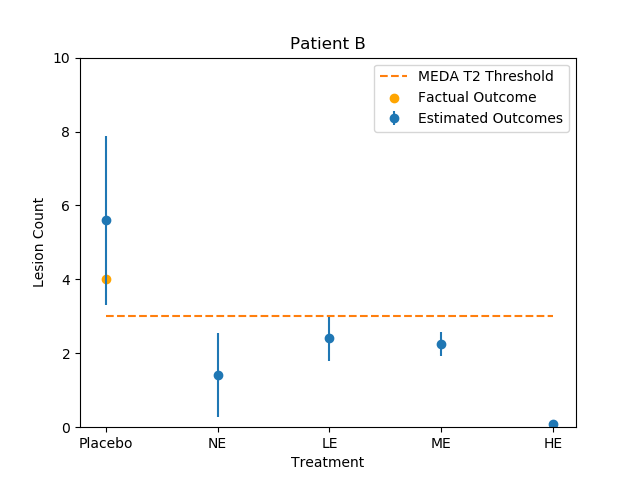} 
}}
\end{figure}

\section{Conclusions}
In this paper, we introduce the first medical imaging-based deep learning model for recommending optimal treatments in MS. The model predicts future NE-T2 counts and MEDA-T2 with high precision on 5 different treatments, and finds sub-groups with heterogeneous treatment effects. However, highly effective suppression of new lesion formation may have only a modest effect on long term disability progression. Current work is focused on predicting stronger markers of disability progression, so as to improve the value of the decision support tool.  Additionally, the model's recommendations have the potential to balance efficacy against treatment associated risks and patient preference. However, our current support tool uses linear scaling of risk between treatments.
A comprehensive risk adjustment model that incorporates patient preferences, side effects, cost and other inconveniences would provide a more holistic clinical support tool but is beyond the scope of this paper.   Future improvements could also be made by estimating treatment effect uncertainty~\cite{JessonCausalFailure} and explicitly optimizing adjusted CATE~\cite{zhao2020mutli-treatment}.
\newpage

\midlacknowledgments{This investigation was supported by the International Progressive Multiple Sclerosis Alliance (award reference number PA-1412-02420), the companies who generously provided the clinical trial data which made it possible: Biogen, BioMS, MedDay, Novartis, Roche / Genentech, and Teva, the Canada Institute for Advanced Research (CIFAR) Artificial Intelligence Chairs program (Arbel), the Natural Sciences and Engineering Research Council of Canada (Arbel), an end MS Personnel Award (Falet) and an AI for MS (Arbel) grant from the Multiple Sclerosis Society of Canada, a Canada Graduate Scholarship-Masters Award from the Canadian Institutes of Health Research (Falet), and the Fonds de recherche Santé / Ministère de la Santé et des Services sociaux training program for specialty medicine residents with an interest in pursuing a research career, Phase 1 (Falet). Supplementary computational resources and technical support were provided by Calcul Québec, WestGrid, and Compute Canada. Additionally, the authors would like to thank Louis Collins and Mahsa Dadar for preprocessing the MRI data, Zografos Caramanos, Alfredo Morales Pinzon, Charles Guttmann and István Mórocz for collating the clinical data, Sridar Narayanan. Maria-Pia Sormani for their MS expertise, and Behrooz Mahasseni for many helpful discussions during model development.}

\bibliography{Durso-Finley22}

\begin{thebibliography}{52}
\providecommand{\natexlab}[1]{#1}
\providecommand{\url}[1]{\texttt{#1}}
\expandafter\ifx\csname urlstyle\endcsname\relax
  \providecommand{\doi}[1]{doi: #1}\else
  \providecommand{\doi}{doi: \begingroup \urlstyle{rm}\Url}\fi

\bibitem[Agarap(2018)]{Relu}
Abien~Fred Agarap.
\newblock Deep learning using rectified linear units (relu).
\newblock \emph{CoRR}, abs/1803.08375, 2018.
\newblock URL \url{http://arxiv.org/abs/1803.08375}.

\bibitem[Ascarza(2018)]{AscarzeQuartileAD}
Eva Ascarza.
\newblock Retention futility: Targeting high-risk customers might be
  ineffective.
\newblock \emph{Journal of Marketing Research}, 55\penalty0 (1):\penalty0
  80--98, 2018.
\newblock \doi{10.1509/jmr.16.0163}.
\newblock URL \url{https://doi.org/10.1509/jmr.16.0163}.

\bibitem[Ching et~al.(2018)Ching, Zhu, and Garmire]{Ching2018CoxnnetAA}
Travers Ching, Xun Zhu, and Lana~X. Garmire.
\newblock Cox-nnet: An artificial neural network method for prognosis
  prediction of high-throughput omics data.
\newblock \emph{PLoS Computational Biology}, 14, 2018.

\bibitem[Collins et~al.(1994)Collins, Neelin, Peters, and
  Evans]{collinsregistration}
D.~L. Collins, P.~Neelin, T.~M. Peters, and A.~Evans.
\newblock {{A}utomatic 3{D} intersubject registration of {M}{R} volumetric data
  in standardized {T}alairach space}.
\newblock \emph{J Comput Assist Tomogr}, 18\penalty0 (2):\penalty0 192--205,
  1994.

\bibitem[Collins and C.~Evans(2011)]{collinsANIMAL}
Louis Collins and A.~C.~Evans.
\newblock Animal: Validation and applications of nonlinear registration-based
  segmentation.
\newblock \emph{International Journal of Pattern Recognition and Artificial
  Intelligence}, 11, 11 2011.
\newblock \doi{10.1142/S0218001497000597}.

\bibitem[Doyle et~al.(2017)Doyle, Precup, Arnold, and
  Arbel]{Doyle2017PredictingFD}
Andrew Doyle, Doina Precup, Douglas~L. Arnold, and Tal Arbel.
\newblock Predicting future disease activity and treatment responders for
  multiple sclerosis patients using a bag-of-lesions brain representation.
\newblock In \emph{MICCAI}, 2017.

\bibitem[Doyle et~al.(2018)Doyle, Elliott, Karimaghaloo, Subbanna, Arnold, and
  Arbel]{DoyleLesionDetection}
Andrew Doyle, Colm Elliott, Zahra Karimaghaloo, Nagesh Subbanna, Douglas~L.
  Arnold, and Tal Arbel.
\newblock Lesion detection, segmentation and prediction in multiple sclerosis
  clinical trials.
\newblock In Alessandro Crimi, Spyridon Bakas, Hugo Kuijf, Bjoern Menze, and
  Mauricio Reyes, editors, \emph{Brainlesion: Glioma, Multiple Sclerosis,
  Stroke and Traumatic Brain Injuries}, pages 15--28, Cham, 2018. Springer
  International Publishing.
\newblock ISBN 978-3-319-75238-9.

\bibitem[Fonov et~al.(2011)Fonov, Evans, Botteron, Almli, McKinstry, Collins,
  Ball, Byars, Schapiro, Bommer, Carr, German, Dunn, Rivkin, Waber, Mulkern,
  Vajapeyam, Chiverton, Davis, Koo, Marmor, Mrakotsky, Robertson, McAnulty,
  Brandt, Fletcher, Kramer, Yang, McCormack, Hebert, Volero, Botteron,
  McKinstry, Warren, Nishino, Almli, Todd, Constantino, McCracken, Levitt,
  Alger, O'Neil, Toga, Asarnow, Fadale, Heinichen, Ireland, Wang, Moss,
  Zimmerman, Bintliff, Bradford, Newman, Evans, Arnaoutelis, Pike, Collins,
  Leonard, Paus, Zijdenbos, Das, Fonov, Fu, Harlap, Leppert, Milovan, Vins,
  Zeffiro, Van~Meter, Lange, Froimowitz, Botteron, Almli, Rainey, Henderson,
  Nishino, Warren, Edwards, Dubois, Smith, Singer, Wilber, Pierpaoli, Basser,
  Chang, Koay, Walker, Freund, Rumsey, Baskir, Stanford, Sirocco, Gwinn-Hardy,
  Spinella, McCracken, Alger, Levitt, and O'Neill]{stxspace}
V.~Fonov, A.~C. Evans, K.~Botteron, C.~R. Almli, R.~C. McKinstry, D.~L.
  Collins, W.~S. Ball, A.~W. Byars, M.~Schapiro, W.~Bommer, A.~Carr, A.~German,
  S.~Dunn, M.~J. Rivkin, D.~Waber, R.~Mulkern, S.~Vajapeyam, A.~Chiverton,
  P.~Davis, J.~Koo, J.~Marmor, C.~Mrakotsky, R.~Robertson, G.~McAnulty, M.~E.
  Brandt, J.~M. Fletcher, L.~A. Kramer, G.~Yang, C.~McCormack, K.~M. Hebert,
  H.~Volero, K.~Botteron, R.~C. McKinstry, W.~Warren, T.~Nishino, C.~R. Almli,
  R.~Todd, J.~Constantino, J.~T. McCracken, J.~Levitt, J.~Alger, J.~O'Neil,
  A.~Toga, R.~Asarnow, D.~Fadale, L.~Heinichen, C.~Ireland, D.~J. Wang,
  E.~Moss, R.~A. Zimmerman, B.~Bintliff, R.~Bradford, J.~Newman, A.~C. Evans,
  R.~Arnaoutelis, G.~B. Pike, D.~L. Collins, G.~Leonard, T.~Paus, A.~Zijdenbos,
  S.~Das, V.~Fonov, L.~Fu, J.~Harlap, I.~Leppert, D.~Milovan, D.~Vins,
  T.~Zeffiro, J.~Van~Meter, N.~Lange, M.~P. Froimowitz, K.~Botteron, C.~R.
  Almli, C.~Rainey, S.~Henderson, T.~Nishino, W.~Warren, J.~L. Edwards,
  D.~Dubois, K.~Smith, T.~Singer, A.~A. Wilber, C.~Pierpaoli, P.~J. Basser,
  L.~C. Chang, C.~G. Koay, L.~Walker, L.~Freund, J.~Rumsey, L.~Baskir,
  L.~Stanford, K.~Sirocco, K.~Gwinn-Hardy, G.~Spinella, J.~T. McCracken, J.~R.
  Alger, J.~Levitt, and J.~O'Neill.
\newblock {{U}nbiased average age-appropriate atlases for pediatric studies}.
\newblock \emph{Neuroimage}, 54\penalty0 (1):\penalty0 313--327, Jan 2011.

\bibitem[Fotso(2018)]{Fotso2018DeepNN}
Stephane Fotso.
\newblock Deep neural networks for survival analysis based on a multi-task
  framework.
\newblock \emph{ArXiv}, abs/1801.05512, 2018.

\bibitem[Freedman et~al.(2020)Freedman, devonshire, Duquette, Giacomini,
  Giuliani, Levin, Montalban, Morrow, Oh, Rotstein, and Yeh]{MSguidelines}
Mark Freedman, Virginia devonshire, Pierre Duquette, Paul Giacomini, Fabrizio
  Giuliani, Michael Levin, Xavier Montalban, Sarah Morrow, Jiwon Oh, Dalia
  Rotstein, and E.~Yeh.
\newblock Treatment optimization in multiple sclerosis: Canadian ms working
  group recommendations.
\newblock \emph{Canadian Journal of Neurological Sciences / Journal Canadien
  des Sciences Neurologiques}, 47:\penalty0 1--76, 04 2020.
\newblock \doi{10.1017/cjn.2020.66}.

\bibitem[González et~al.(2018)González, Ash, Vegas-Sánchez-Ferrero,
  Onieva~Onieva, Rahaghi, Ross, Díaz, San José~Estépar, and
  Washko]{SmokingPrognosis}
G.~González, S.~Y. Ash, G.~Vegas-Sánchez-Ferrero, J.~Onieva~Onieva, F.~N.
  Rahaghi, J.~C. Ross, A.~Díaz, R.~San José~Estépar, and G.~R. Washko.
\newblock {{D}isease {S}taging and {P}rognosis in {S}mokers {U}sing {D}eep
  {L}earning in {C}hest {C}omputed {T}omography}.
\newblock \emph{Am J Respir Crit Care Med}, 197\penalty0 (2):\penalty0
  193--203, 01 2018.

\bibitem[Guelman(2015)]{PropensityScoreUsage}
Leandro~Axel Guelman.
\newblock Optimal personalized treatment learning models with insurance
  applications.
\newblock 2015.

\bibitem[Gutierrez and Gérardy(2017)]{Gutierrez2017}
Pierre Gutierrez and Jean-Yves Gérardy.
\newblock Causal inference and uplift modelling: A review of the literature.
\newblock volume~67, pages 1--13. PMLR, 12 2017.
\newblock URL \url{https://proceedings.mlr.press/v67/gutierrez17a.html}.

\bibitem[Ha et~al.(2018)Ha, Chin, Karcich, Liu, Chang, Mutasa, Sant, Wynn,
  Connolly, and Jambawalikar]{TreatmentEffectBreastTumor}
Richard Ha, Christine Chin, Jenika Karcich, Michael Liu, Peter Chang, Simukayi
  Mutasa, Eduardo Sant, Ralph Wynn, Eileen Connolly, and Sachin Jambawalikar.
\newblock Prior to initiation of chemotherapy, can we predict breast tumor
  response? deep learning convolutional neural networks approach using a breast
  mri tumor dataset.
\newblock \emph{Journal of Digital Imaging}, 32, 10 2018.
\newblock \doi{10.1007/s10278-018-0144-1}.

\bibitem[Hauser et~al.(2017)Hauser, Bar-Or, Comi, Giovannoni, Hartung, Hemmer,
  Lublin, Montalban, Rammohan, Selmaj, Traboulsee, Wolinsky, Arnold,
  Klingelschmitt, Masterman, Fontoura, Belachew, Chin, Mairon, Garren, and
  Kappos]{OPERA}
Stephen~L. Hauser, Amit Bar-Or, Giancarlo Comi, Gavin Giovannoni, Hans-Peter
  Hartung, Bernhard Hemmer, Fred Lublin, Xavier Montalban, Kottil~W. Rammohan,
  Krzysztof Selmaj, Anthony Traboulsee, Jerry~S. Wolinsky, Douglas~L. Arnold,
  Gaelle Klingelschmitt, Donna Masterman, Paulo Fontoura, Shibeshih Belachew,
  Peter Chin, Nicole Mairon, Hideki Garren, and Ludwig Kappos.
\newblock Ocrelizumab versus interferon beta-1a in relapsing multiple
  sclerosis.
\newblock \emph{New England Journal of Medicine}, 376\penalty0 (3):\penalty0
  221--234, 2017.
\newblock \doi{10.1056/NEJMoa1601277}.
\newblock URL \url{https://doi.org/10.1056/NEJMoa1601277}.
\newblock PMID: 28002679.

\bibitem[Havrdova et~al.(2013)Havrdova, Hutchinson, Kurukulasuriya, Raghupathi,
  Sweetser, Dawson, and Gold]{DEFINE}
E.~Havrdova, M.~Hutchinson, N.~C. Kurukulasuriya, K.~Raghupathi, M.~T.
  Sweetser, K.~T. Dawson, and R.~Gold.
\newblock {{O}ral {B}{G}-12 (dimethyl fumarate) for relapsing-remitting
  multiple sclerosis: a review of {D}{E}{F}{I}{N}{E} and {C}{O}{N}{F}{I}{R}{M}.
  {E}valuation of: {G}old {R}, {K}appos {L}, {A}rnold {D}, et al.
  {P}lacebo-controlled phase 3 study of oral {B}{G}-12 for relapsing multiple
  sclerosis. {N} {E}ngl {J} {M}ed 2012;367:1098-107; and {F}ox {R}{J}, {M}iller
  {D}{H}, {P}hillips {J}{T}, et al. {P}lacebo-controlled phase 3 study of oral
  {B}{G}-12 or glatiramer in multiple sclerosis. {N} {E}ngl {J} {M}ed
  2012;367:1087-97}.
\newblock \emph{Expert Opin Pharmacother}, 14\penalty0 (15):\penalty0
  2145--2156, Oct 2013.

\bibitem[He et~al.(2015)He, Zhang, Ren, and Sun]{he2015deepResNet}
Kaiming He, Xiangyu Zhang, Shaoqing Ren, and Jian Sun.
\newblock Deep residual learning for image recognition, 2015.

\bibitem[Jaroszewicz(2014)]{Jaroszewicz2014UpliftMW}
Szymon Jaroszewicz.
\newblock Uplift modeling with survival data.
\newblock 2014.

\bibitem[Jesson et~al.(2020)Jesson, Mindermann, Shalit, and
  Gal]{JessonCausalFailure}
Andrew Jesson, S{\"{o}}ren Mindermann, Uri Shalit, and Yarin Gal.
\newblock Identifying causal effect inference failure with uncertainty-aware
  models.
\newblock \emph{CoRR}, abs/2007.00163, 2020.
\newblock URL \url{https://arxiv.org/abs/2007.00163}.

\bibitem[Katzman et~al.(2018)Katzman, Shaham, Cloninger, Bates, Jiang, and
  Kluger]{DeepSurv}
Jared~L. Katzman, Uri Shaham, Alexander Cloninger, Jonathan Bates, Tingting
  Jiang, and Yuval Kluger.
\newblock Deepsurv: personalized treatment recommender system using a cox
  proportional hazards deep neural network.
\newblock \emph{BMC Medical Research Methodology}, 18\penalty0 (1), Feb 2018.
\newblock ISSN 1471-2288.
\newblock \doi{10.1186/s12874-018-0482-1}.
\newblock URL \url{http://dx.doi.org/10.1186/s12874-018-0482-1}.

\bibitem[Krstajic et~al.(2014)Krstajic, Buturovic, Leahy, and
  Thomas]{NestedCrossVal}
Damjan Krstajic, Ljubomir Buturovic, David Leahy, and Simon Thomas.
\newblock Cross-validation pitfalls when selecting and assessing regression and
  classification models.
\newblock \emph{Journal of cheminformatics}, 6:\penalty0 10, 03 2014.
\newblock \doi{10.1186/1758-2946-6-10}.

\bibitem[Kurtzke(1983)]{Kurtzke1444EDSS}
John~F. Kurtzke.
\newblock Rating neurologic impairment in multiple sclerosis.
\newblock \emph{Neurology}, 33\penalty0 (11):\penalty0 1444--1444, 1983.
\newblock ISSN 0028-3878.
\newblock \doi{10.1212/WNL.33.11.1444}.
\newblock URL \url{https://n.neurology.org/content/33/11/1444}.

\bibitem[Le~Page and Edan(2018)]{msescalationguidelines}
E.~Le~Page and G.~Edan.
\newblock {{I}nduction or escalation therapy for patients with multiple
  sclerosis?}
\newblock \emph{Rev Neurol (Paris)}, 174\penalty0 (6):\penalty0 449--457, Jun
  2018.

\bibitem[Lin et~al.(2018)Lin, Tong, Gao, Guo, Du, Yang, Guo, Xiao, Du, Qu, and
  ~]{AlzeheimersPrognosis}
Weiming Lin, Tong Tong, Qinquan Gao, Di~Guo, Xiaofeng Du, Yonggui Yang, Gang
  Guo, Min Xiao, Min Du, Xiaobo Qu, and The Alzheimer’s Disease
  Neuroimaging~Initiative ~.
\newblock Convolutional neural networks-based mri image analysis for the
  alzheimer’s disease prediction from mild cognitive impairment.
\newblock \emph{Frontiers in Neuroscience}, 12:\penalty0 777, 2018.
\newblock ISSN 1662-453X.
\newblock \doi{10.3389/fnins.2018.00777}.
\newblock URL
  \url{https://www.frontiersin.org/article/10.3389/fnins.2018.00777}.

\bibitem[Loshchilov and Hutter(2019)]{loshchilov2019AdamW}
Ilya Loshchilov and Frank Hutter.
\newblock Decoupled weight decay regularization, 2019.

\bibitem[Louizos et~al.(2017)Louizos, Shalit, Mooij, Sontag, Zemel, and
  Welling]{louizos2017causal}
Christos Louizos, Uri Shalit, Joris Mooij, David Sontag, Richard Zemel, and Max
  Welling.
\newblock Causal effect inference with deep latent-variable models, 2017.

\bibitem[Maas(2013)]{Maas2013RectifierNI}
Andrew~L. Maas.
\newblock Rectifier nonlinearities improve neural network acoustic models.
\newblock 2013.

\bibitem[Manjón et~al.(2010)Manjón, Coupé, Martí-Bonmatí, Collins, and
  Robles]{Denoising}
J.~V. Manjón, P.~Coupé, L.~Martí-Bonmatí, D.~L. Collins, and M.~Robles.
\newblock {{A}daptive non-local means denoising of {M}{R} images with spatially
  varying noise levels}.
\newblock \emph{J Magn Reson Imaging}, 31\penalty0 (1):\penalty0 192--203, Jan
  2010.

\bibitem[Nair et~al.(2020)Nair, Precup, Arnold, and Arbel]{TanyaUncMSseg}
Tanya Nair, Doina Precup, Douglas~L. Arnold, and Tal Arbel.
\newblock Exploring uncertainty measures in deep networks for multiple
  sclerosis lesion detection and segmentation.
\newblock \emph{Medical Image Analysis}, 59:\penalty0 101557, 2020.
\newblock ISSN 1361-8415.
\newblock \doi{https://doi.org/10.1016/j.media.2019.101557}.
\newblock URL
  \url{https://www.sciencedirect.com/science/article/pii/S1361841519300994}.

\bibitem[Nichyporuk et~al.(2021)Nichyporuk, Cardinell, Szeto, Mehta, Tsaftaris,
  Arnold, and Arbel]{nichyporuk2021cohort}
Brennan Nichyporuk, Jillian Cardinell, Justin Szeto, Raghav Mehta, Sotirios
  Tsaftaris, Douglas~L Arnold, and Tal Arbel.
\newblock {C}hort {B}ias {A}daptation in {A}ggregated {D}atasets for {L}esion
  {S}egmentation.
\newblock In \emph{Domain Adaptation and Representation Transfer, and
  Affordable Healthcare and AI for Resource Diverse Global Health}, pages
  101--111. Springer, 2021.

\bibitem[Nielsen et~al.(2018)Nielsen, Hansen, Tietze, and
  Mouridsen]{StrokePrognosis}
A.~Nielsen, M.~B. Hansen, A.~Tietze, and K.~Mouridsen.
\newblock {{P}rediction of {T}issue {O}utcome and {A}ssessment of {T}reatment
  {E}ffect in {A}cute {I}schemic {S}troke {U}sing {D}eep {L}earning}.
\newblock \emph{Stroke}, 49\penalty0 (6):\penalty0 1394--1401, 06 2018.

\bibitem[Prosperini et~al.(2009)Prosperini, Gallo, Petsas, Borriello, and
  Pozzilli]{OneYearMSINF-B}
Luca Prosperini, Valentina Gallo, Nikolaos Petsas, Giovanna Borriello, and
  C~Pozzilli.
\newblock One-year mri scan predicts clinical response to interferon beta in
  multiple sclerosis.
\newblock \emph{European journal of neurology : the official journal of the
  European Federation of Neurological Societies}, 16:\penalty0 1202--9, 06
  2009.
\newblock \doi{10.1111/j.1468-1331.2009.02708.x}.

\bibitem[Roy et~al.(2018)Roy, Butman, Reich, Calabresi, and
  Pham]{MSSEGFULLYCONV}
Snehashis Roy, John~A. Butman, Daniel~S. Reich, Peter~A. Calabresi, and
  Dzung~L. Pham.
\newblock Multiple sclerosis lesion segmentation from brain {MRI} via fully
  convolutional neural networks.
\newblock \emph{CoRR}, abs/1803.09172, 2018.
\newblock URL \url{http://arxiv.org/abs/1803.09172}.

\bibitem[Rubin(1974)]{Rubin1974EstimatingCE}
Donald~B. Rubin.
\newblock Estimating causal effects of treatments in randomized and
  nonrandomized studies.
\newblock \emph{Journal of Educational Psychology}, 66:\penalty0 688--701,
  1974.

\bibitem[Rudick et~al.(2006)Rudick, Lee, Simon, and Fisher]{RRMSactivity}
Richard Rudick, Jar-Chi Lee, Jack Simon, and Elizabeth Fisher.
\newblock Significance of t2 lesions in multiple sclerosis: A 13-year
  longitudinal study.
\newblock \emph{Annals of neurology}, 60:\penalty0 236--42, 08 2006.
\newblock \doi{10.1002/ana.20883}.

\bibitem[Río et~al.(2008)Río, Rovira, Tintoré, Huerga, Nos, Tellez, Tur,
  Comabella, and Montalban]{RioScore}
J~Río, À~Rovira, M~Tintoré, E~Huerga, C~Nos, N~Tellez, C~Tur, M~Comabella,
  and X~Montalban.
\newblock Relationship between mri lesion activity and response to ifn-beta in
  relapsing–remitting multiple sclerosis patients.
\newblock \emph{Multiple Sclerosis Journal}, 14\penalty0 (4):\penalty0
  479--484, 2008.
\newblock \doi{10.1177/1352458507085555}.
\newblock PMID: 18562504.

\bibitem[Sepahvand et~al.(2020)Sepahvand, Arnold, and Arbel]{nazactivitypred}
Nazanin Sepahvand, Douglas Arnold, and Tal Arbel.
\newblock Cnn detection of new and enlarging multiple sclerosis lesions from
  longitudinal mri using subtraction images.
\newblock pages 127--130, 04 2020.
\newblock \doi{10.1109/ISBI45749.2020.9098554}.

\bibitem[Sepahvand et~al.(2019)Sepahvand, Hassner, Arnold, and
  Arbel]{NazFirstActivityPaper}
Nazanin~Mohammadi Sepahvand, Tal Hassner, Douglas~L. Arnold, and Tal Arbel.
\newblock Cnn prediction of future disease activity for multiple sclerosis
  patients from baseline mri and lesion labels.
\newblock In Alessandro Crimi, Spyridon Bakas, Hugo Kuijf, Farahani Keyvan,
  Mauricio Reyes, and Theo van Walsum, editors, \emph{Brainlesion: Glioma,
  Multiple Sclerosis, Stroke and Traumatic Brain Injuries}, pages 57--69, Cham,
  2019. Springer International Publishing.
\newblock ISBN 978-3-030-11723-8.

\bibitem[Shalit et~al.(2017)Shalit, Johansson, and Sontag]{shalit2017TARNET}
Uri Shalit, Fredrik~D. Johansson, and David Sontag.
\newblock Estimating individual treatment effect: generalization bounds and
  algorithms, 2017.

\bibitem[Shi et~al.(2019)Shi, Blei, and Veitch]{shi2019adaptingDRAGONNET}
Claudia Shi, David~M. Blei, and Victor Veitch.
\newblock Adapting neural networks for the estimation of treatment effects,
  2019.

\bibitem[Sled et~al.(2002)Sled, Zijdenbos, and
  Evans]{intensitynonuniformcorrection}
J.G. Sled, Alex Zijdenbos, and Alan Evans.
\newblock A nonparametric method for automatic correction of intensity
  nonuniformity in mri data. i.
\newblock \emph{E.E.E. Transactions on Medical Imaging}, 17:\penalty0 87--97,
  01 2002.

\bibitem[Soltys et~al.(2014)Soltys, Jaroszewicz, and
  Rzepakowski]{Soltys2014EnsembleMF}
Michal Soltys, Szymon Jaroszewicz, and Piotr Rzepakowski.
\newblock Ensemble methods for uplift modeling.
\newblock \emph{Data Mining and Knowledge Discovery}, 29:\penalty0 1531--1559,
  2014.

\bibitem[Sormani et~al.(2013)Sormani, Rio, Tintorè, Signori, Li, Cornelisse,
  Stubinski, Stromillo, Montalban, and Stefano]{MPScoringCoxPH}
MP~Sormani, J~Rio, M~Tintorè, A~Signori, D~Li, P~Cornelisse, B~Stubinski,
  ML~Stromillo, X~Montalban, and N~De Stefano.
\newblock Scoring treatment response in patients with relapsing multiple
  sclerosis.
\newblock \emph{Multiple Sclerosis Journal}, 19\penalty0 (5):\penalty0
  605--612, 2013.
\newblock \doi{10.1177/1352458512460605}.
\newblock URL \url{https://doi.org/10.1177/1352458512460605}.
\newblock PMID: 23012253.

\bibitem[Srivastava et~al.(2014)Srivastava, Hinton, Krizhevsky, Sutskever, and
  Salakhutdinov]{JMLR:v15:Dropout}
Nitish Srivastava, Geoffrey Hinton, Alex Krizhevsky, Ilya Sutskever, and Ruslan
  Salakhutdinov.
\newblock Dropout: A simple way to prevent neural networks from overfitting.
\newblock \emph{Journal of Machine Learning Research}, 15\penalty0
  (56):\penalty0 1929--1958, 2014.
\newblock URL \url{http://jmlr.org/papers/v15/srivastava14a.html}.

\bibitem[Sun et~al.(2019)Sun, Zhang, Chen, and Luo]{BRATSSurvivalprog}
Li~Sun, Songtao Zhang, Hang Chen, and Lin Luo.
\newblock Brain tumor segmentation and survival prediction using multimodal mri
  scans with deep learning.
\newblock \emph{Frontiers in Neuroscience}, 13:\penalty0 810, 2019.
\newblock ISSN 1662-453X.
\newblock \doi{10.3389/fnins.2019.00810}.
\newblock URL
  \url{https://www.frontiersin.org/article/10.3389/fnins.2019.00810}.

\bibitem[Tousignant et~al.(2019)Tousignant, Lema\^itre, Precup, Arnold, and
  Arbel]{pmlr-v102-tousignant19a}
Adrian Tousignant, Paul Lema\^itre, Doina Precup, Douglas~L. Arnold, and Tal
  Arbel.
\newblock Prediction of disease progression in multiple sclerosis patients
  using deep learning analysis of mri data.
\newblock In M.~Jorge Cardoso, Aasa Feragen, Ben Glocker, Ender Konukoglu, Ipek
  Oguz, Gozde Unal, and Tom Vercauteren, editors, \emph{Proceedings of The 2nd
  International Conference on Medical Imaging with Deep Learning}, volume 102
  of \emph{Proceedings of Machine Learning Research}, pages 483--492. PMLR,
  08--10 Jul 2019.
\newblock URL \url{https://proceedings.mlr.press/v102/tousignant19a.html}.

\bibitem[Ulyanov et~al.(2017)Ulyanov, Vedaldi, and
  Lempitsky]{ulyanov2017instance}
Dmitry Ulyanov, Andrea Vedaldi, and Victor Lempitsky.
\newblock Instance normalization: The missing ingredient for fast stylization,
  2017.

\bibitem[Valverde et~al.(2017)Valverde, Cabezas, Roura, González-Villà,
  Pareto, Vilanova, Ramió-Torrentà, Àlex Rovira, Oliver, and
  Lladó]{VALVERDE2017159MSseg}
Sergi Valverde, Mariano Cabezas, Eloy Roura, Sandra González-Villà, Deborah
  Pareto, Joan~C. Vilanova, Lluís Ramió-Torrentà, Àlex Rovira, Arnau
  Oliver, and Xavier Lladó.
\newblock Improving automated multiple sclerosis lesion segmentation with a
  cascaded 3d convolutional neural network approach.
\newblock \emph{NeuroImage}, 155:\penalty0 159--168, 2017.
\newblock ISSN 1053-8119.
\newblock \doi{https://doi.org/10.1016/j.neuroimage.2017.04.034}.
\newblock URL
  \url{https://www.sciencedirect.com/science/article/pii/S1053811917303270}.

\bibitem[Vollmer et~al.(2014)Vollmer, Sorensen, Selmaj, Zipp, Havrdova, Cohen,
  Sasson, Gilgun-Sherki, and Arnold]{BRAVO}
T.~L. Vollmer, P.~S. Sorensen, K.~Selmaj, F.~Zipp, E.~Havrdova, J.~A. Cohen,
  N.~Sasson, Y.~Gilgun-Sherki, and D.~L. Arnold.
\newblock {{A} randomized placebo-controlled phase {I}{I}{I} trial of oral
  laquinimod for multiple sclerosis}.
\newblock \emph{J Neurol}, 261\penalty0 (4):\penalty0 773--783, Apr 2014.

\bibitem[Xu et~al.(2019)Xu, Hosny, Zeleznik, Parmar, Coroller, Franco, Mak, and
  Aerts]{XuLung58}
Yiwen Xu, Ahmed Hosny, Roman Zeleznik, Chintan Parmar, Thibaud Coroller, Idalid
  Franco, Raymond~H. Mak, and Hugo~J.W.L. Aerts.
\newblock Deep learning predicts lung cancer treatment response from serial
  medical imaging.
\newblock \emph{Clinical Cancer Research}, 25\penalty0 (11):\penalty0
  3266--3275, 2019.
\newblock ISSN 1078-0432.
\newblock \doi{10.1158/1078-0432.CCR-18-2495}.
\newblock URL \url{https://clincancerres.aacrjournals.org/content/25/11/3266}.

\bibitem[Zhao et~al.(2017)Zhao, Fang, and Simchi{-}Levi]{ZhaoCTS}
Yan Zhao, Xiao Fang, and David Simchi{-}Levi.
\newblock Uplift modeling with multiple treatments and general response types.
\newblock \emph{CoRR}, abs/1705.08492, 2017.
\newblock URL \url{http://arxiv.org/abs/1705.08492}.

\bibitem[Zhao and Harinen(2020)]{zhao2020mutli-treatment}
Zhenyu Zhao and Totte Harinen.
\newblock Uplift modeling for multiple treatments with cost optimization, 2020.

\end{thebibliography}

\appendix

\newpage
\section{Implementation Details}
\label{sec:ImpDets}
The MRI sequences are first clipped between $+/-3$ standard deviations and then normalized to $N(0,1)$ per sequence. The MRI sequences are then resampled to 2x2x2 resolution and cropped for a final dimension of 72x76x52. The clinical data is normalized to $N(0,1)$. 
\par As mentioned in the Network Architecture section, the trunk of the model consists of three ResNet blocks followed by several MLPs. Each ResNet block contains two convolutional blocks followed by a residual addition. Each convolutional block contains a convolution (kernel size 3, stride 1), Instance Normalization~\cite{ulyanov2017instance}, a dropout layer~\cite{JMLR:v15:Dropout} with $p=0.3$, and a LeakyReLU activation~\cite{Maas2013RectifierNI}. Each ResNet block, with the exception of the last, is followed by an max pooling operation with kernel size 2. In the three ResNet blocks, the number of kernels for each convolution is [32, 64, 128] respectively. After the three ResNet blocks, the latents are flattened using a global average pool before concatenating the features with the clinical information and inputting the combined latent space to the MLPs. Each of the 5 MLPs in the network consist of three hidden layers which have dimensions [128,32,16] and use ReLU activations \cite{Relu} with no dropout. For training, we used the AdamW optimizer\cite{loshchilov2019AdamW} with a learning rate of .0001 and a batch size of 8.

For models using imaging data and clinical data, the clinical data included age, gender and baseline EDSS. For the models using clinical data only, the clinical data included age, gender, baseline EDSS, baseline T2 lesion volume, and baseline Gad lesion count.

\newpage
\section{Lesion Counts}
\setcounter{figure}{0}
\renewcommand{\thefigure}{B.\arabic{figure}}
\label{sec:Lesion Counts}
\begin{figure}[!h]
    \floatconts
    {fig:ptp}
    {\caption{Future NE-T2 Lesion Count Histogram.}}
    {\centering
        \includegraphics[width=0.5\textwidth]{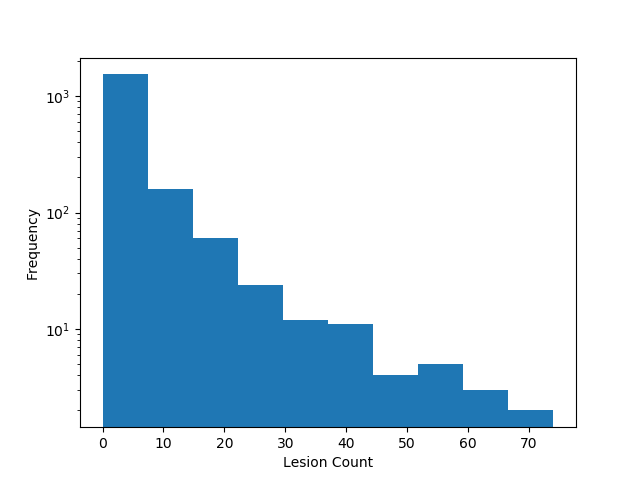} 
    }
\end{figure}

\begin{figure}[!h]
    \floatconts
    {fig:countAD_all}
    {\caption{Future NE-T2 Lesion Counts by Treatment}}
    
    {
    \subfigure{
    \centering
    \includegraphics[width=.45\linewidth]{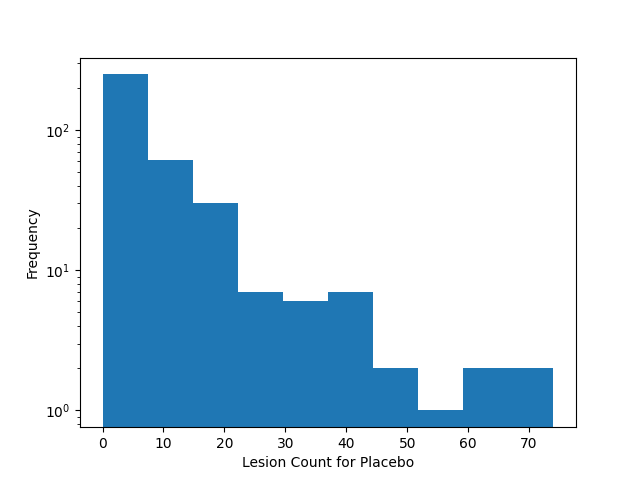}}
    \subfigure{
    \centering
    \includegraphics[width=.45\linewidth]{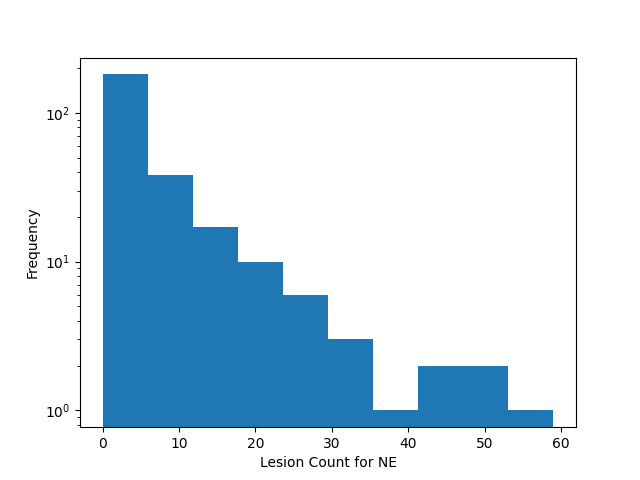}}
    \\
    \subfigure{
    \centering
    \includegraphics[width=.45\linewidth]{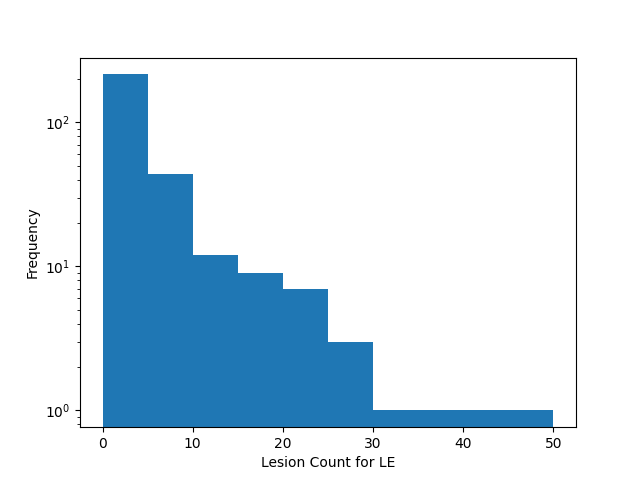}}
      \subfigure{
      \centering
    \includegraphics[width=.45\linewidth]{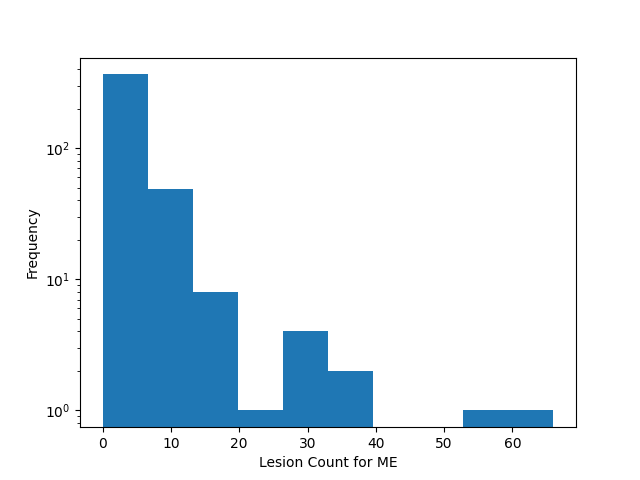}}
    \\
    \subfigure{
    \includegraphics[width=.45\linewidth]{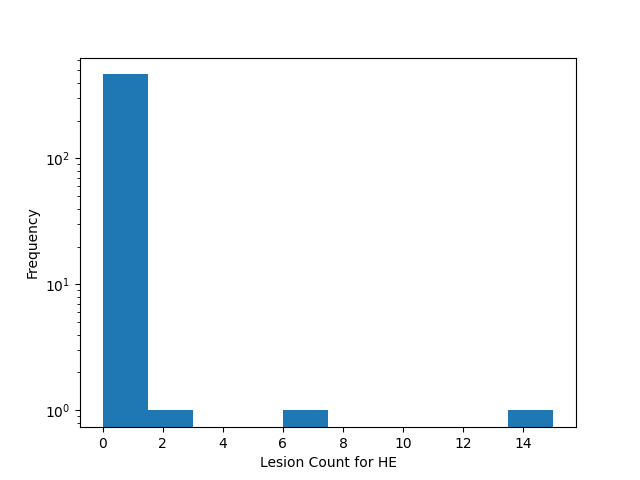}}
    }
\end{figure}

\newpage
\section{Treatment Effect Analysis with the binary MEDA-T2 outcome}
\setcounter{figure}{0}
\renewcommand{\thefigure}{C.\arabic{figure}}
\begin{figure}[!h]
\floatconts
  {fig:ADNEWT2_binary}
  {\caption{Treatment Effect Analysis. (a) Average difference in frequency of MEDA-T2 between treatment-placebo pairs, binned according to tertiles of predicted treatment effect size. (b) Frequency of risk-adjusted MEDA-T2 for individuals who did (blue) or did not (orange) receive the recommended treatment, compared to random treatment assignment (green). Incremental risk values ($\lambda$) are varied on the $x$-axis.}}
  {
    \subfigure{
    \centering
    \label{subfig:countAD_binaryboth}
    \includegraphics[width=.45\linewidth]{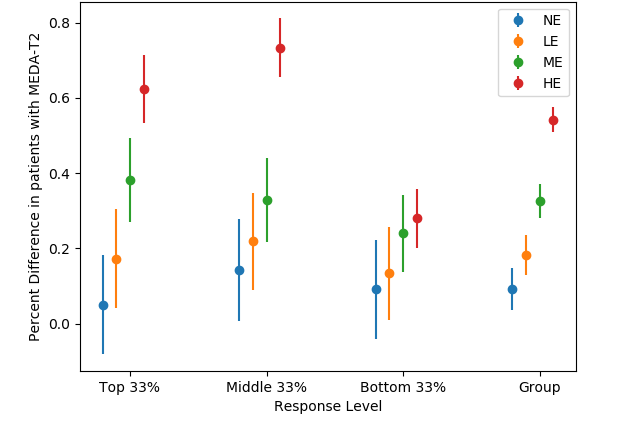}}
      \subfigure{
      \centering
    \includegraphics[width=.45\linewidth]{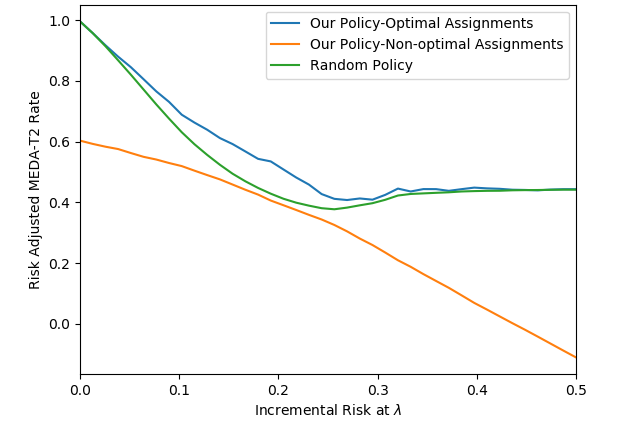}}}
\end{figure}
\newpage
\section{Additional Uplift Bins}
\label{sec:ADquindec}
\setcounter{figure}{0}
\renewcommand{\thefigure}{D.\arabic{figure}}
\begin{figure}[!h]
\floatconts
  {fig:ADNEWT2_count_510}
  {\caption{Average difference in NE-T2 lesion count between treatment-placebo pairs, binned according to quintiles (a) and deciles (b) of predicted treatment effect size.}}
  {
    \subfigure{
    \centering
    \label{subfig:countAD_5}
    \includegraphics[width=.45\linewidth]{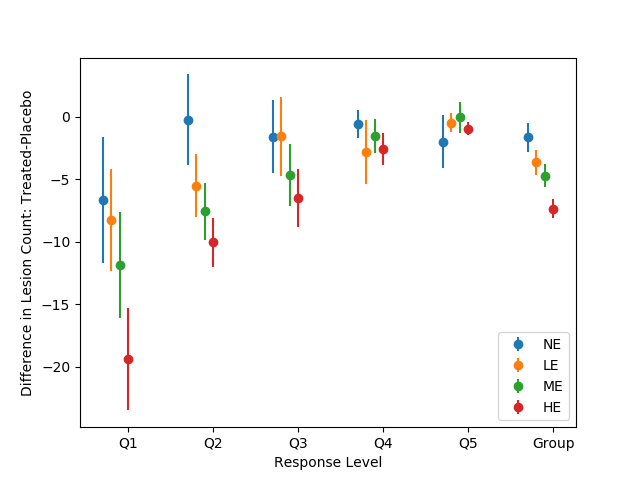}}
      \subfigure{
      \centering
    \includegraphics[width=.45\linewidth]{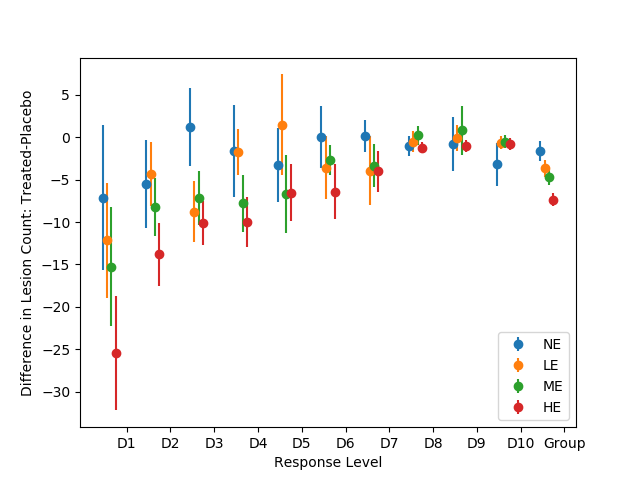}}}
\end{figure}
\newpage
\section{Additional Results}
\label{sec:AddRes}
\setcounter{table}{0}
\renewcommand{\thetable}{E.\arabic{table}}
\begin{table}[htbp]
    \floatconts
    {tab:factualROCAUC}
    {\caption{ROC-AUC for the binary MEDA-T2 outcome.}}
     {\resizebox{1\columnwidth}{!}{\begin{tabular}{|c|c|c|c|c|c|c|c|}
     \hline
     Model Type & $+$ Clinical & Multi-Head & Placebo  &  NE &  LE &   ME  &  HE  \\
     \hline
     Baseline & & & 0.5 &  0.5 & 0.5 & 0.5 & 0.5 \\
     \hline
     Clinical Only& \checkmark &  & 0.76 +\- .03 & 0.719+/-.02 & 0.745+/-.05  &  0.73+/-.03 & 0.46+/- .07 \\    
     \hline
     Binary Classification& \checkmark &  & 0.77 +\- .03 & 0.69+/-.05 & 0.68+/-.06  &  0.759+/-.03 & 0.5 +/- .11 \\ 
     \hline
     Binary Classification & \checkmark & \checkmark &  0.818 +/- .01 & 0.738 +/- .071 & 0.770 +/- .001 & 0.753 +/- .014 & 0.488 +/- .0017 \\
     \hline
     Binary  Classification && \checkmark & 0.772 +/- .04 &  0.682 +/- .04 & 0.73 +/- .01 & 0.751+/- .04 & 0.497+/- .04\\
     \hline
     Binarized Regression & \checkmark & \checkmark & \textbf {0.836 +/- .01} & \textbf {0.749 +/- .0021} & \textbf {0.783 +/- .001 }& \textbf {0.769 +/- .014} & 0.488 +/- .0017\\
     \hline
\end{tabular}}}
\end{table}
\begin{table}[htbp]
    \floatconts
    {tab:factualRegressionTargetMAE}
    {\caption{MAE for log lesion count regression against baseline}}
     {\resizebox{.8\columnwidth}{!}{\begin{tabular}{|c|c|c|c|c|c|}
     \hline
      Model &  Placebo  &  NE &  LE &   ME  &  HE  \\
      \hline
      Baseline & 0.94 &  0.98 & 0.89 & 0.789 & 0.072\\
      \hline
       MAE  & 0.658 +/- .08 &  0.839 +/- .059 & 0.70 +/-.052 & 0.64 +/- .07 & 0.07+/- .01 \\
      \hline
\end{tabular}}
}
\end{table}
\section{Pretrial Patient Statistics.}
\label{sec:PretrialStats}
\setcounter{table}{0}
\renewcommand{\thetable}{F.\arabic{table}}

\begin{table}[htbp]
    \floatconts
    {tab:PretrialStats}
    {\caption{Baseline clinical and scalar MRI metrics for our dataset. Standard deviations are in parentheses.}}
     {\resizebox{1\columnwidth}{!}{\begin{tabular}{|c|c|c|c|c|c|c|c|c|}
     \hline
     
     Trial/Treatment &  BRAVO/Placebo & DEFINE/Placebo & BRAVO/NE &  BRAVO/LE  & OPERA 1/ME & OPERA 2/ME  & OPERA 1/HE & OPERA 2/HE \\
     \hline
     N &  278 & 94 &  261 & 295  & 223 & 208 & 236 & 232   \\
     \hline
     Age   & 37.95 (9.27) & 37.8 (9.51) &  37.03 (9.2) & 38.29 (9.45) & 37.2 (9.25) & 37.3 (8.95) & 37.1 (9.27) & 37.5 (8.85)   \\
     \hline
     Gender(Male Fraction) & 0.29 & 0.25 &  0.29 & 0.31 & 0.33 & 0.34  & 0.34 & 0.37  \\
     \hline
     Baseline EDSS & 2.71 (1.16) & 2.46 (1.23) &  2.67 (1.23) & 2.64 (1.14) & 2.7 (1.27) & 2.68 (1.37) & 2.77 (1.21) & 2.68 (1.27)   \\   
     \hline
     T2 Lesion Volume  & 7.82 (8.714) & 6.67 (8.2) &  9.28 (9.8) & 8.4 (9.2)  & 9.28 (11.1) & 10.0 (12.3) &  10.96 (14.21) & 10.83 (14.25)  \\
     \hline
     Gad Count  & 1.12 (3.24) & 1.84 (3.91) &  1.61 (4.40) & 1.48 (3.5) & 1.535 (4.75) & 1.87 (4.47) &  1.73 (4.35) & 1.85 (4.8)  \\
     \hline
\end{tabular}}}
\end{table}

\begin{table}[htbp]
    \floatconts
    {tab:TreatmentDistributions}
    {\caption{Treatments used for the model by trial.}}
    {\resizebox{.8\columnwidth}{!}{\begin{tabular}{|c|c|c|c|c|c|}
        \hline
         Trial & High Efficacy Treatment & Moderate Efficacy Treatment & Lower Efficacy Treatment & No Efficacy Treatment & Placebo  \\
         \hline
         OPERA 1 & Ocrelizumab & INFb-1a SC & & &  \\
         \hline
         OPERA 2 & Ocrelizumab & INFb-1a SC & & &  \\
         \hline
         BRAVO &  & & Avonex & Laquinimod &
         Placebo   \\
         \hline
         DEFINE &  &  & & & Placebo  \\
         \hline
    \end{tabular}}}
\end{table}
\newpage

\section{Significance Values}
\label{sec:SignificanceValues}
\setcounter{table}{0}
\renewcommand{\thetable}{G.\arabic{table}}
\begin{table}[htbp]
    \floatconts
    {tab:Significance Values}
    {\caption{P values for group differences between response groups for the regression task shown in figure \figureref{subfig:countAD}. Column headers indicate the two responder groups (tertiles) that are being compared.}}
     {\resizebox{.8\columnwidth}{!}{\begin{tabular}{|c|c|c|c|c|}
     \hline
     Grouping & NE & LE & ME & HE \\ 
     \hline
     Top 33\%-Middle 33\% & .0043 & $<$.001 & $<$.001 & $<$.001  \\
          \hline
     Top 33\%-Bottom 33\% & .0028 &  $<$.001 & $<$.001 & $<$.001  \\ 
          \hline
     Top 33\% -Group 33\% & .031 & $<$.001 & $<$.001 & $<$.001   \\ 
          \hline
     Middle 33\%-Bottom 33\% & .89 & .03 & $<$.001 & $<$.001  \\ 
          \hline
     Middle 33\%-Group 33\%  & .70 &  .19 & .97 & .43   \\ 
          \hline
     Bottom 33\%-Group 33\%  & .78 &  $<$.001 &  $<$.001 & $<$.001   \\
     \hline
\end{tabular}}}
\end{table}

\section{MRI Preprocessing}
 Scans were first denoised \cite{Denoising}, corrected for intensity heterogeneity  \cite{intensitynonuniformcorrection}, and normalized into the range 0-100. Second, for each patient, the T2w, PD, and FLAIR scans were co-registered to the structural T1w scan using a 6-parameter rigid registration and a mutual information objective function~\cite{collinsregistration}. The T1w scans were then registered to an average template defining stereotaxic space \cite{collinsANIMAL, stxspace}.
All volumes are resampled onto a 1 mm isotropic grid using the T1-to-stx space transformation (for the T1w data) or the transformation that results from concatenating the contrast-to-T1 and T1-to-stx transformation (for the other contrasts).
\end{document}